# Decoupling Many-Body Interactions in CeO$_2$ (111) Oxygen Vacancy Structure: Insights from Machine-Learning and Cluster Expansion

Yujing Zhang[1,2], Zhong-Kang Han[3], Beien Zhu[2,4], Xiaojuan Hu[5], Maria Troppenz[6], Santiago Rigamonti[6], Hui Li[1]*, Claudia Draxl[6], M. Verónica Ganduglia-Pirovano,[7]* Yi Gao[2,4,8]*

[1]Beijing Advanced Innovation Center for Soft Matter Science and Engineering, Beijing University of Chemical Technology, Beijing 100029, China

[2]Key Laboratory of Interfacial Physics and Technology, Shanghai Institute of Applied Physics, Chinese Academy of Sciences, Shanghai 201800, China

[3]School of Materials Science and Engineering, Zhejiang University, Hangzhou,310027, China

[4] Phonon Science Research Center for Carbon Dioxide, Shanghai Advanced Research Institute, Chinese Academy of Sciences, Shanghai 201210, China

[5]Fritz-Haber-Institut der Max-Planck-Gesellschaft, Faradayweg 4-6, 14195 Berlin, Germany

[6]Institut für Physik und Iris Adlershof, Humboldt-Universität zu Berlin, Zum Großen Windkanal 2, 12489 Berlin, Germany

[7]Instituto de Catálisis y Petroleoquímica of the Consejo Superior de Investigaciones Científicas, 28049, Madrid, Spain

[8]Key Laboratory of Low-Carbon Conversion Science & Engineering, Shanghai Advanced Research Institute, Chinese Academy of Sciences, Shanghai 201210, China

Email: hli@buct.edu.cn; vgp@icp.csic.es; gaoyi@sari.ac.cn

**Abstract:** Oxygen vacancies ($V_O$'s) are of paramount importance in influencing the properties and applications of ceria ($CeO_2$). Yet, comprehending the distribution and nature of the $V_O$'s poses a significant challenge due to the vast number of electronic configurations and intricate many-body interactions among $V_O$'s and polarons ($Ce^{3+}$'s). In this study, we employed a combination of LASSO regression in machine learning, in conjunction with a cluster expansion model and first-principles calculations to decouple the interactions among the $Ce^{3+}$'s and $V_O$'s, thereby circumventing the limitations associated with sampling electronic configurations. By separating these interactions, we identified specific electronic configurations characterized by the most favorable $V_O$-$Ce^{3+}$ attractions and the least $Ce^{3+}$-$Ce^{3+}$/$V_O$-$V_O$ repulsions, which are crucial in determining the stability of vacancy structures. Through more than $10^8$ Metropolis Monte Carlo samplings of Vo's and $Ce^{3+}$ in the near-surface of CeO$_2$(111), we explored potential configurations within an 8×8 supercell. Our findings revealed that oxygen vacancies tend to aggregate and are most abundant in the third oxygen layer, primarily due to extensive geometric relaxation-an aspect previously overlooked. This behavior is notably dependent on the concentration of Vo. This work introduces a novel theoretical framework for unraveling the complex vacancy structures in metal oxides, with potential applications in redox and catalytic chemistry.

## INTRODUCTION

Cerium oxide ($CeO_2$), known for its high concentration of surface oxygen vacancies ($V_O$'s), exhibits remarkable oxygen storage capacity and redox catalytic performance, finding applications in diverse areas such as automobile exhaust gas treatment, hydrogen production, purification, and fuel cells. [1-10] The $CeO_2$ (111) crystal facet, known for its stability and accessibility, has been extensively studied [11-19]. Experiments involving scanning tunneling microscopy (STM) and atomic resolution dynamic force microscopy (DFM) have provided insights into the various near-surface Vo structures at the $CeO_2$(111) surface [20, 21]. The theoretical interpretation of the distributions of these vacancies has proven to be a challenge [22-30]. Early studies identified the formation of a single $V_O$, leaving two electrons localized on the 4f orbitals of two $Ce^{4+}$ ions, converting them into $Ce^{3+}$ ions [22]. Li et al.[23] and Ganduglia-Pirovano et al. [24,25] suggested that the energetically favored location for an isolated oxygen vacancy is in the subsurface layer, with the excess electrons localized in two next nearest neighbors cations (NNN). This configuration was found to be approximately 0.2 eV more stable than the NNN of the surface vacancy and significantly lower in energy than the $V_O$ in deeper oxygen layers. Further research explained the observed high stability of an ordered (2×2) oxygen subsurface vacancy structure at a $V_O$ concentration of 1/4 [21,25]. However, precise sampling of the $V_O$'s structures remains challenging due to the varied stability introduced by $Ce^{3+}$ ions at different coordination spheres around the $V_O$'s [31]. As the number of Vo's increases, the number of $Ce^{3+}$ also rises, leading to an exponential increase in possible electronic configurations. The spatial correlation among $Ce^{3+}$'s, $V_O$'s, and the mutual spatial correlation between $Ce^{3+}$'s and $V_O$'s become exceedingly complex and are closely tied to the stability of configurations. Understanding the nature of $V_O$'s structures and making accurate predictions about $V_O/Ce^{3+}$ configurations requires a precise evaluation of the stabilities of numerous electronic configurations, considering interactions among $Ce^{3+}$-$Ce^{3+}$, $V_O$-$Ce^{3+}$, $V_O$-$V_O$, and their couplings. However, this level of complexity surpasses the present computational capacity.

In this study, we employed a combination of machine learning (ML) methods, the cluster expansion (CE) model, and first-principles calculations to decouple the interactions among $Ce^{3+}$'s and $V_O$'s. This approach allowed us to sample electronic configurations of $V_O$'s at the $CeO_2$(111) surface more effec-tively. We identified specific configurations characterized by maximum $V_O$-$Ce^{3+}$ attractions and minimum $Ce^{3+}$-$Ce^{3+}$/$V_O$-$V_O$ repulsions. Notably, as the concentration of $V_O$'s increases, they tend to distribute toward the third oxygen layer rather than concentrating near the surface, as observed with isolated $V_O$'s. This unique behavior is attributed to a distinct geometric relaxation, a factor that was overlooked in earlier studies, which primarily considered lower oxygen vacancy concentration and/or focused on the surface and subsurface layers [23-25,32]. The integration of machine learning and computational models has significantly enhanced our understanding of the intricate interactions and configurations of oxygen vacancies ($V_O$'s) and $Ce^{3+}$'s species on the $CeO_2$(111) surface. It's evident that the stability of oxygen vacancies is crucial in comprehending the surface physics and chemistry of reducible oxides, suggesting a potential need for reinterpreting earlier data. Given that the surface chemistry of such oxides is largely influenced by defects, there is ample opportunity for unexpected discoveries.

## RESULTS and DISCUSSION

**Decoupling of the many-body interaction by a machine-learning model**

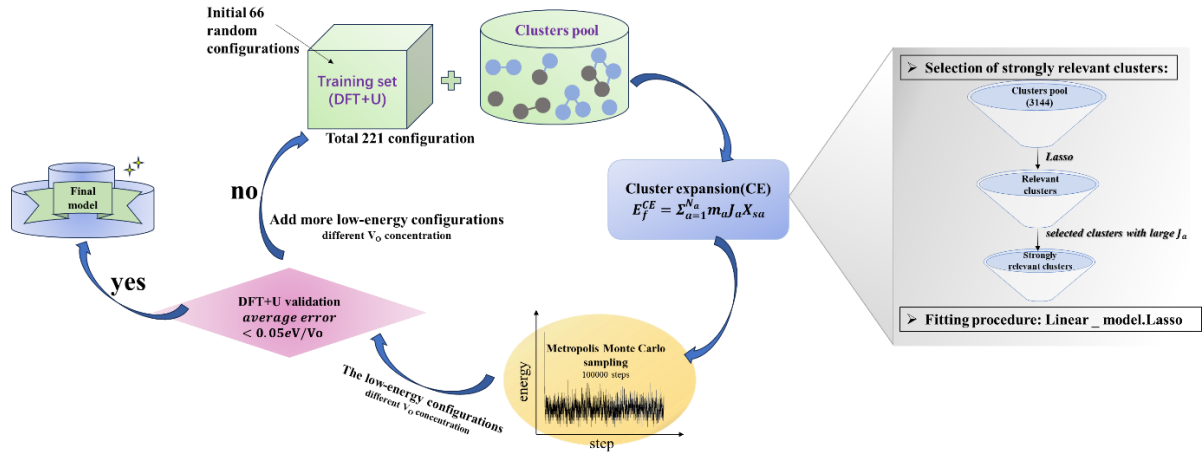

**Scheme 1. Machine learning flowchart of CE model.**

To determine the $V_O$ formation energy ($E_f^{CE}$) of any configuration of $V_O$'s and their associated $Ce^{3+}$'s, we constructed a CE model that accounts for interactions among $Ce^{3+}$-$Ce^{3+}$, $V_O$-$Ce^{3+}$, $V_O$-$V_O$, and their couplings (clusters). As shown in **Scheme 1**, the training data for our CE model consisted of DFT+U calculations involving diverse random configurations of $V_O$'s structures. These structures encompassed $V_O$ concentrations Θ of 1/16, 1/8, 3/16, and 1/4. These $V_O$ concentrations pertain to the monolayer oxygen atoms and are defined as Θ= $n$/$N$, where $N$=16 and $n$=1, 2, 3, 4. Here, $N$ represents the number of oxygen atoms in a non-reduced oxygen layer within the surface unit cell, and $n$ is the number of $V_O$'s in the unit cell. Due to the fact that the formation energies of bulk $V_O$'s are much larger than those near the surface, as indicated by previous studies,[24,33-34] and our preliminary tests (**Supplementary Figure 1**), we have confined the Vo's to the three topmost oxygen layers (the surface, subsurface, and the third oxygen atom layer). Similarly, $Ce^{3+}$ ions are confined to the two topmost cationic layers (the surface and subsurface). Our objective is to investigate whether oxygen vacancies will remain near the surface, as hypothesized, or will unexpectedly occupy sites in the third layer upon increasing the oxygen vacancy concentration. Subsequently, the $V_O$'s formation energies

$$E_f^{DFT} = E_{CeO_{2-x}}^{relax} + \frac{n}{2} E_{O_2} - E_{CeO_2}^{pristine} \qquad (1)$$

$E_{CeO_{2-x}}^{relax}$ is the energy of configuration with $n$ $V_O$'s and relaxed geometry, $E_{O_2}$ and $E_{CeO_2}^{pristine}$ are the total energy of the isolated $O_2$ molecule and the pristine slab were calculated. In the DFT+U calculation process, the oxidation state of a given Ce atom can be estimated by evaluating its local magnetic moment, defined as the difference between up and down spin on the atoms. This can be estimated by integrating the site- and angular momentum projected spin-resolved density of states over spheres with radii chosen as the Wigner−Seitz radii of the PAW potentials. In the case of reduced Ce ions, where the occupation of Ce f states is close to 1 and the magnetic moment is ~1 $\mu_B$, these ions are typically referred to as $Ce^{3+}$. **Supplementary Figure 2** illustrates isosurfaces of the difference between spin up and spin down charges for the example of four subsurface oxygen vacancies and eight $Ce^{3+}$ ions. The localization of the charge in f-like orbitals is clearly observable. Additionally, $Ce^{3+}$-O bond lengths (~2.49 Å) are larger than those of $Ce^{4+}$-O bonds (~2.37Å).

All clusters consisting of no more than 4 points (where a point represents either a $Ce^{3+}$ or a Vo) were considered, with the largest distance between points (lattice sites) set at 8 Å for two-body, three-body and four-body interactions. The initial number of clusters is 3144. These clusters account for the positions of $V_O$'s and $Ce^{3+}$'s, and the distances between them. The key clusters were selected based on the formation energies of charge-balanced configurations obtained from DFT calculations, which inherently account for all the electronic interactions, including electrostatic interactions. Thus, charge balance is implicit in our model. These clusters serve as parameters within the CE model rather than as independent physical structures. To represent a complete and realistic configuration, multiple clusters must be combined (as shown in **Supplementary Figure 3**). To disentangle the many-body interactions and identify the predominant features with the largest effective cluster interactions, we employed the least absolute shrinkage and selection operator (LASSO) approach[35-37]. LASSO is an effective algorithm to avoid overfitting and select the optimal cluster set. In this approach, the objective function minimized is:

$$\eta = \sum_{i}^{m} \left(E_f^{CE}(J_a)_i - E_f^{DFT}{}_i\right)^2 + \lambda \|J_a\|_1 \tag{2}$$

Here, the first term represents the summation of square error between the predicted and calculated formation energies, and the second term, the $l_1$ regularization term, is the summation of $|J_a|$. The hyperparameter λ is obtained by minimizing a leave-one-out cross-validation error. The leave-one-out cross-validation method was employed to avoid overfitting and evaluate the predictive power of the CE model. From the clusters identifies by LASSO, we isolated those with significant effective cluster interactions ($J_a$) to establish a preliminary model. Subsequently, we performed metropolis Monte Carlo (MC) simulations coupled with machine learning predictions at 1000 K over $10^5$ steps. At each step, the model provided energy predictions, enabling the identification of the 20 lowest-energy configurations across various $V_O$ concentrations (4 concentrations with 5 configurations per concentration). We then verified whether the average prediction error of the Vo formation energy, $E_f^{CE}$, was below 0.05 eV/Vo using DFT+U calculations. If the error exceeded this threshold, new configurations sampled by MC were incorporated into the training set, and the training process continued. The final model converged to standard reference training error metrics (RMSE, MAE), cross-validation error (CV_RMSE, CV_MAE), and DFT verification thresholds for the 20 low-energy configurations. Following 8 iterations of active training, wherein clusters were re-selected with the pool containing 3144 clusters at each iteration, we identified 15 key clusters from the initial pool of 3144 clusters by comparing errors across different clusters and energy deviation of the lowest-energy configurations. This comparison is presented in **Supplementary Table 1** and **Supplementary Table 2**. In total, 221 configurations were utilized in the final training process of our CE model. The configurations for training, along with the diversity and uniformity of our training set are shown in **Supplementary Figure 4.**

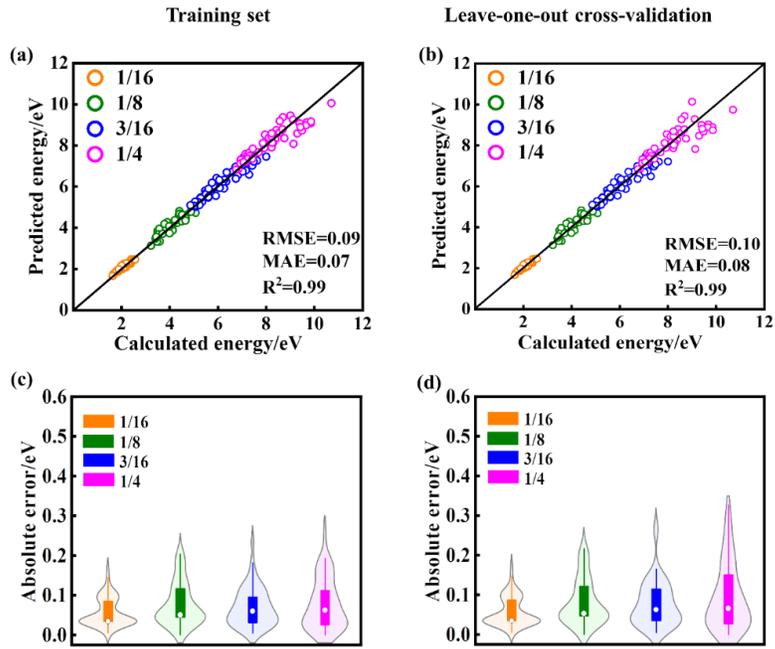

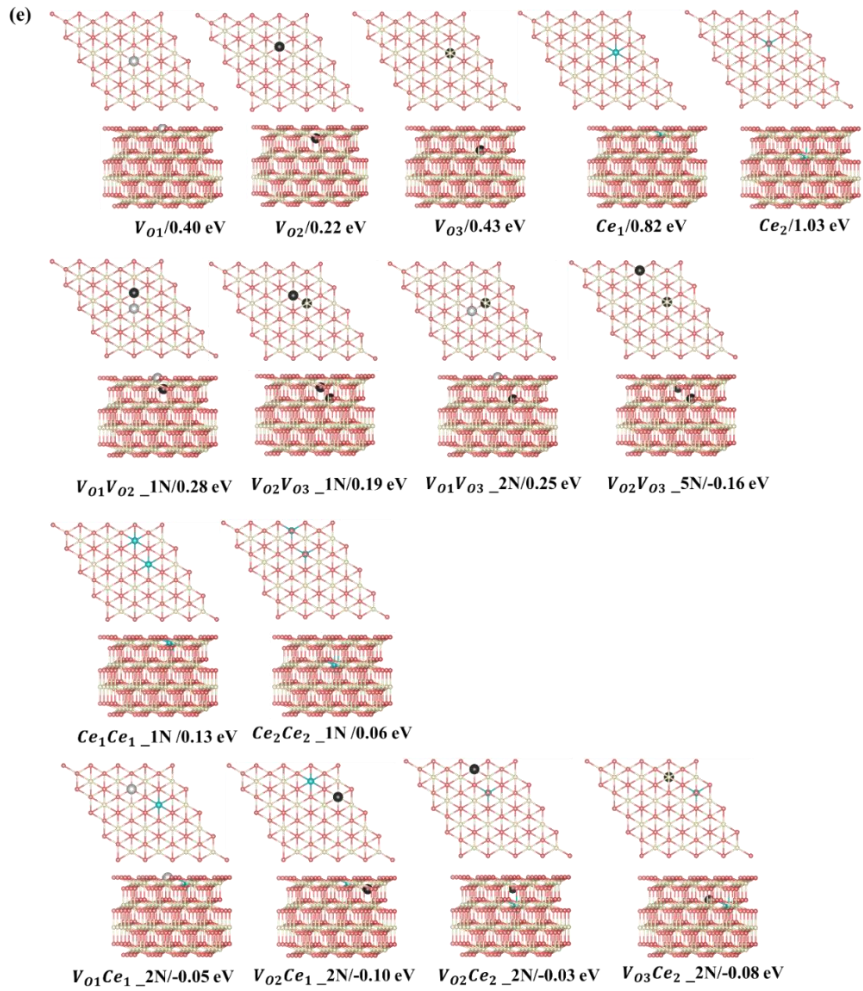

**Fig. 1 Model evaluation and primary features from Machine-Learning.** The error between the cluster-expansion model predicted energy and the DFT energy for **(a)** training set and **(b)** leave-one-out cross-validation. Violin plots of the absolute error in the energy of configurations for **(c)** training set and **(d)** leave-one-out cross-validation. The upper and lower limits of the rectangles represent the 75th and 25th percentiles of the distribution, the internal white dot marks the median (50th percentile), and the upper and lower limits of the thin line extending from the rectangle bars indicate the minimum and maximum errors. The violin area represents the probability of data distribution. **(e)** 15 primary features predicted by machine learning and their energies, with structures depicted in the unrelaxed state. $Ce^{4+}$ ions, $Ce^{3+}$ ions, oxygen atoms, and surface oxygen vacancy are depicted in white, blue, pink, and light gray respectively. Subsurface and third oxygen layer vacancy are depicted in black.

The training maximum-absolute error (MAE) is 0.07 eV/$V_O$ (**Fig.1(a)**) and the cross-validation MAE is 0.08 eV/$V_O$ (**Fig.1(b)**). The detailed distributions of the absolute errors between the CE predicted formation energy and the DFT calculated formation energy of Vo's are displayed in **Fig.1(c)** and **Fig.1(d)**. The error distributions are concentrated below 0.1 eV, as indicated by the median line and the broad area of the violin plot. The largest deviations between the calculated and predicted energies for the training set and leave-one-out cross-validation arises from the high-energy unstable configurations. The formation energies of the 15 primary cluster features identified by decoupling the many-body interaction are shown in **Fig.1(e)**. In this notation, $V_{On}Ce_{m\_}lN$, $n$ and $m$ denote the oxygen atomic layer and the Ce atomic layer where the $V_O$ and $Ce^{3+}$ are located, respectively. The $l$ denotes the positional relationship definition of $V_O$-$V_O$, $V_O$-$Ce^{3+}$, $Ce^{3+}$-$Ce^{3+}$ in the order of nearest neighbors (as shown in **Supplementary Figure 5** and **Supplementary Table 3**). The monomer configuration reveals that the isolated vacancy prefers the subsurface ($V_{O2}$ = 0.22 eV), which is approximately 0.2 eV more stable than the surface vacancy ($V_{O1}$ = 0.40 eV) and the 3$^{rd}$ layer vacancy ($V_{O3}$ = 0.43 eV). In contrast, $Ce^{3+}$ ion prefers the surface ($Ce_1$ = 0.82 eV) over the subsurface ($Ce_2$ = 1.03 eV), which is in line with previous knowledge[23]. Regarding two-body interactions, there is a clear repulsion between the $V_O$'s in nearest-neighbor positions in different layers ($V_{O1}V_{O2}\_1N$, $V_{O2}V_{O3}\_1N$, $V_{O1}V_{O3}\_2N$). Among these interactions, $V_{O1}V_{O2}\_1N$ exhibits the strongest repulsion (0.28 eV), while $V_{O2}V_{O3}\_1N$ is the weakest (0.19 eV). However, Vo's in the subsurface and the 3$^{rd}$ layers have a strong attraction when forming a specific pattern ($V_{O2}V_{O3}\_5N$ = −0.16 eV). For cerium ions, surface $Ce^{3+}$ ions in nearest neighbor positions exhibit stronger repulsion ($Ce_1Ce_1\_1N$ = 0.13 eV) than subsurface $Ce^{3+}$ ions ($Ce_2Ce_2\_1N$ = 0.06 eV). Additionally, the next-nearest neighbor interaction between Vo and $Ce^{3+}$ ($V_{On}Ce_{n}\_2N$) is consistently attractive, in line with previous studies suggesting that $Ce^{3+}$ ions prefer the NNN position to $V_O$'s [24]. Specifically, for surface $Ce^{3+}$ ions, the strongest attraction is with the formation of a NNN configuration with subsur-face $V_O$'s ($V_{O2}Ce_1\_2N = -0.10$ eV), while for the subsurface $Ce^{3+}$, the strongest attractions are with 3rd layer $V_O$'s ($V_{O3}Ce_2\_2N = -0.08$ eV). The contributions of other cluster features are negligible.

**Machine-learning model predicts stability of configurations**

The special motifs described above can be used to understand the specific stability of the local configurations, serving as building blocks in comprehending the stability of complete physical configurations. Leveraging on the 15 primary features obtained through decoupling, one can readily predict the stability of any vacancy structure concerning the positions of $V_O$'s and $Ce^{3+}$'s. **Supplementary Figure 6** gives some motifs of single $V_O$ with two $Ce^{3+}$'s, Among these configurations, the subsurface $V_O$ with two NNN surface $Ce^{3+}$'s is most stable, being 0.28 eV and 0.42 eV lower in energy than the surface and 3rd layer Vo with two NNN surface Ce3+ ions, respectively. This finding is consistent with the previous studies[23,25]. However, in the case of multi-vacancies, the situation becomes more complicated, as shown in **Fig.2**. For the surface $Ce^{3+}$ ions **(a-e)**, as the positions of $V_O$'s and the relative position between the $V_O$'s and the $Ce^{3+}$'s ions vary, it becomes evident that the motif **(a)** involves the most stable subsurface $V_{O2}$ and exhibits the most attractive interaction between $Ce^{3+}$ and $V_O$ ($V_{O2}Ce_1\_2N$) compared to other configurations. The distance between the $V_O$'s is that of 3rd neighbors (7.72 Å) in the oxygen layer, consistent with previous work [25]. On the other hand, for subsurface $Ce^{3+}$ **(f-j)**, the motif **(f)**, which satisfies both strong $V_{O3}Ce_2\_2N$ and $V_{O2}V_{O3}\_5N$ attractions, is most stable. This indicated that the 3rd layer - $V_O$'s are stable when subsurface $Ce^{3+}$ are present. In summary, specific local configurations with high stability have been identified.

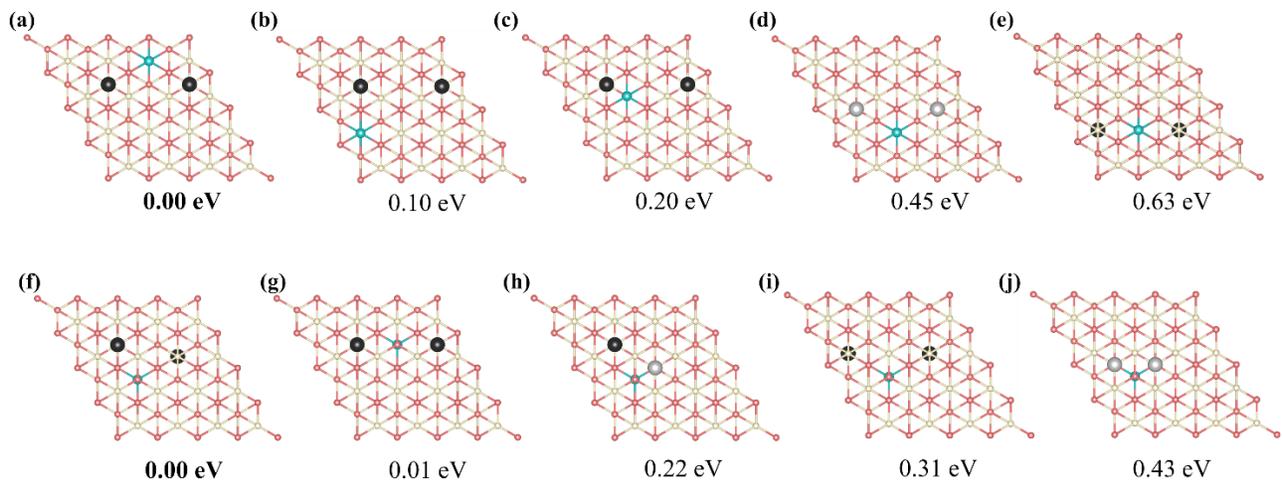

**Fig. 2 Predicted relative energies of selected motifs of Vo's and Ce³⁺'s based on the identified 15 primary features. (a)-(e)** $Ce^{3+}$ ion is located in the first cerium layer, **(f)-(j)** $Ce^{3+}$ ion is located in the

second cerium layer. $Ce^{4+}$ion, $Ce^{3+}$ion, oxygen atoms, and surface oxygen vacancy are depicted in white, blue, pink, and light gray respectively, subsurface and third oxygen layer vacancy are depicted in black. Structures are depicted in the unrelaxed state.

Based on the analysis above, we can understand the lowest-energy configurations with different concentrations (Θ = 1/16, 1/8, 3/16, and 1/4) using a (4×4) supercell obtained from the Scheme 1 protocol. As shown in **Supplementary Figure 7**, these low-energy configurations were predicted by the CE model and subsequently verified through DFT+U calculations. For low $V_O$ concentrations (Θ = 1/16 and 1/8), consistent with previous literature [23,25], subsurface $V_O$'s are the most stable, with excess electrons localized on nearest-neighboring (NNN) cerium positions to the vacancies. Specifically, for Θ = 1/8, in the most stable configuration, Vo's form third-nearest-neighbor pairs in the oxygen plane, in line with previous knowledge [25] (**Supplementary Figure 7**). However, as the $V_O$ concentration increases (Θ = 3/16), vacancies begin to appear in the 3$^{rd}$ oxygen layer in addition to the subsurface layer, rather than in the surface layer. A configuration featuring two $V_O$'s in the subsurface and one $V_O$ in 3$^{rd}$ oxygen layer (cf. $V_{O2}V_{O3}\_5N$ in **Fig.1**) becomes more stable than the configuration in which all $V_O$'s are located in the subsurface (**Supplementary Figure 7**). With a further increase in $V_O$ concentration to 1/4, more $V_O$'s distribute to the 3$^{rd}$ layer. A stable structure emerges, composed of $V_{O2}V_{O3}\_5N$ and $V_{O2}Ce_1\_2N$ features (**Fig.3(a)**). In this structure, termed $A_{2233}$, two $V_O$'s are in the subsurface layer, while two $V_O$'s are in the 3$^{rd}$ oxygen layer, termed as $A_{2233}$. The $Ce^{3+}$ ions are positioned in NNN positions relative to the vacancies. This configuration is energetically favorable by 0.09 eV compared to the proposed (2×2) structure from previous literature, where all $V_O$'s are in the subsurface and are third-nearest neighbors in the oxygen plane (**Fig.3(b)**) [25]. It has been suggested that vacancy-induced lattice relaxations effects play an essential role in determining the spacing between $V_O$'s in the (2×2) structure. Moreover, although the outermost cerium layer is usually the energy-preferred location for $Ce^{3+}$ ions, in both of these structures, they prefer to be in the deeper layer rather than adjacent to a vacancy. In the novel structure (**Fig.3(a)**), two $V_O$'s are located in the third oxygen layer rather than all in the subsurface, with a distance of 6.11Å ($V_{O2}V_{O3}\_5N$ = 6.11Å) between $V_{O2}$ and $V_{O3}$.

**Stability analysis based on DFT calculations**

To gain a deeper understanding of the molecular mechanism behind the stability of $V_O$'s in the 3$^{rd}$ layer rather than the surface layer at high vacancy concentrations, we focus on the $A_{2233}$, $A_{2222}$ and $A_{1122}$ configurations at Θ = 1/4 ($A_{mnpq}$, m, n, p, q denote the oxygen layers where the $V_O$'s are located) (**Fig.3**).

In our analysis we propose that the vacancy formation energy can be decomposed into two components[24]: bond-breaking energy

$$E_b = E_{CeO_{2-x}}^{unrelax} + \frac{n}{2}E_{O_2} - E_{CeO_2}^{pristine} \qquad (3)$$

$E_{CeO_{2-x}}^{unrelax}$ is the energy of a configuration with n V$_O$'s and unrelaxed geometry, $E_{O_2}$ and $E_{CeO_2}^{pristine}$ are the total energy of the isolated O$_2$ molecule and the pristine slab and relaxation energy

$$E_r = E_{CeO_{2-x}}^{relax} - E_{CeO_{2-x}}^{unrelax} \qquad (4)$$

$E_{CeO_{2-x}}^{relax}$ is the energy of configuration with $n$ V$_O$'s and relaxed geometry, as shown in **Table 1**. The relaxation energies for A$_{2233}$, A$_{2222}$, and A$_{1122}$ are −9.85 eV, −8.80 eV, −6.90 eV, respectively.

**Table 1. Vacancy formation energy ($E_f^{DFT}$), relaxation energy ($E_r$), and bond-breaking energy ($E_b$) of configurations with four V$_O$'s (Θ = 1/4). $E_r^*$, $E_b^*$ is sum of the corresponding energies to isolated single V$_O$'s in the A$_{mnpq}$ configurations.**

|  | $E_f^{DFT}$ (eV) | $E_r$ (eV) | $E_b$ (eV) | $E_r^*$ (eV) | $E_b^*$ (eV) |
| --- | --- | --- | --- | --- | --- |
| A$_{2233}$ | 6.75 | −9.85 | 16.60 | −9.88 | 17.28 |
| A$_{2222}$ | 6.84 | −8.80 | 15.64 | −9.8 | 16.32 |
| A$_{1122}$ | 8.04 | −6.90 | 14.94 | −8.98 | 15.90 |
| A$_{3333}$ | 8.20 | −9.53 | 17.77 | −9.96 | 18.24 |

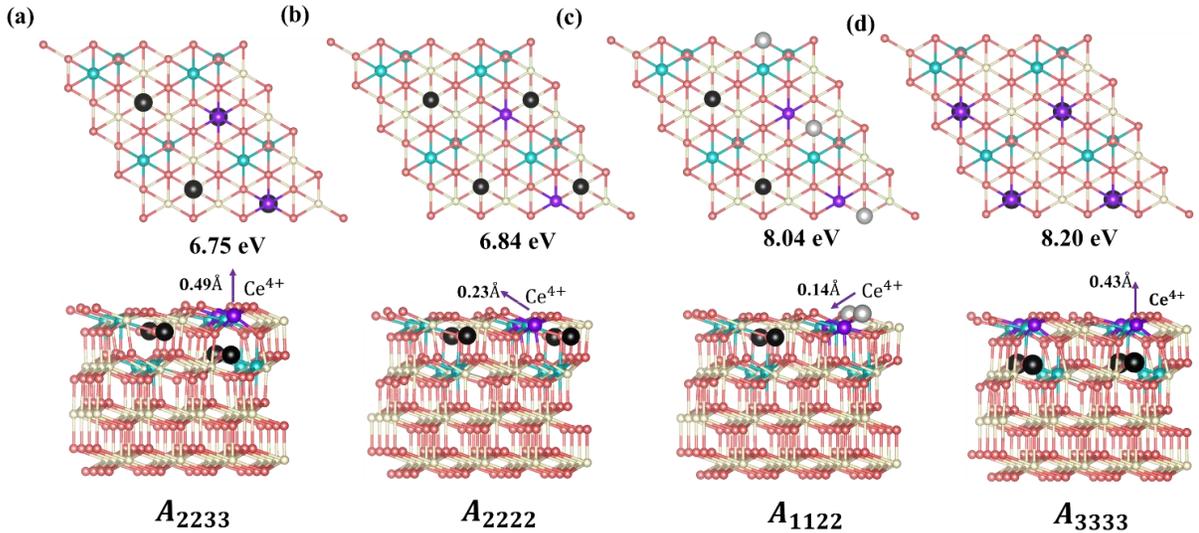

**Fig. 3 Top views, side views, and the Vo's formation energies of the configurations $A_{2233}$, $A_{2222}$, $A_{1122}$ and $A_{3333}$.** (a) configuration $A_{2233}$, with two $V_O$'s in the subsurface layer and two $V_O$'s in the 3rd oxygen layer, (b) configuration $A_{2222}$, with four $V_O$'s in the subsurface layer, (c) configuration $A_{1122}$, with two $V_O$'s in the subsurface layer and two $V_O$'s in the surface layer. (d) configuration $A_{3333}$, with four $V_O$'s in the 3rd layer. $Ce^{4+}$ ions, $Ce^{3+}$ ions, oxygen atoms and surface oxygen vacancy are depicted in white, blue, pink, and light gray respectively, subsurface and third oxygen layer vacancy are depicted in black, purple atoms are $Ce^{4+}$ ions. Top and side views are depicted in the unrelaxed and relaxed state, respectively. Arrows indicate the relaxation direction of the purple-colored $Ce^{4+}$ ions at the surface layer.

When compared to $E_b$, it is evident that the larger $E_r$, resulting from substantial geometric relaxation, is the primary driver behind the stability of the $A_{2233}$ configuration. To further elucidate the impact of the relaxation, we also calculated $E_r$ and $E_b$ for isolated $V_O$'s (Θ = 1/16) and considered the sum of the corresponding energies to isolated single $V_O$'s in the $A_{mnpq}$ configurations. As shown in **Supplementary Table 4** and **Supplementary Figure 8**, the deeper the $V_O$ is located, the more significant $E_r$ and $E_b$ become. When comparing an isolated subsurface $V_O$ to a surface $V_O$, the relaxation energy gain increases (ΔE_r = 0.41 eV) considerably more than bond-breaking energies (ΔE_b = 0.21 eV). This indicates that structure relaxation is necessary for subsurface vacancy stabilization, in line with a previous study [24]. Conversely, when the isolated $V_O$ is positioned in the 3rd layer, $E_b$ is significantly enhanced (0.69 eV) compared to the surface $V_O$, surpassing the increase in relaxation energy (0.45 eV). As a result, a single $V_O$ is favored to be distributed in the subsurface. In summary, there exists a competitive relationship between $E_r$ and $E_b$ [24], where extensive geometric relaxation plays a pivotal role in determining the stability of $V_O$'s at different locations. For configurations $A_{mnpq}$ with four $V_O$'s, the $E_b$ for $A_{2233}$, $A_{2222}$ and $A_{1122}$ (**Table 1**) is approximately 1 eV lower than $E_b^*$ (sum of the corresponding energies of isolated single $V_O$). Furthermore, the relaxation energy gains ($|E_r|$) are reduced compared to isolated single $V_O$'s ($|E_r^*|$) due counteracting vacancy-induced relaxation effects with increasing coverage (from 1/16 to 1/4). However, the relaxation in configuration $A_{2233}$ with 3rd layer $V_O$'s is only minimally impeded (0.03 eV) compared to $A_{2222}$ (1 eV) and $A_{1122}$ (2.08 eV), which contributes to the stability of $A_{2233}$. The stronger relaxation in $A_{2233}$ is reflected in the atomic displacements of surface $Ce^{4+}$ ions nearest to the $V_O$'s. This distance is much larger for $A_{2233}$ (0.49 Å) in average (purple atoms in **Fig.3**) compared to those for $A_{2222}$ (0.23 Å) and $A_{1122}$ (0.14 Å). It is because $Ce^{4+}$ ions nearest neighbor to the $V_O$'s tend to move away from $V_O$ site [25]; $A_{2233}$ provides more space for the upward relaxation of the 6-fold coordinated surface $Ce^{4+}$ ions due to the absence of a Ce–O bond at the opposite position, which is a distinct feature compared to $A_{2222}$ and $A_{1122}$ (**Fig.3**). However, it is not the case that larger the relaxation energy always leads to more stable the corresponding configuration. Based on our CE model, we sampled configurations with four vacancies in the third layer and predicted the stable configuration $A_{3333}$, which was then verified by DFT (Fig. 3). Note that in all four configurations, $Ce^{3+}$ ions are in next-nearest sites to the vacancies. Our analysis revealed that $A_{2233}$ remains the most stable configuration, while $A_{3333}$ is the least stable. Specifically, the $E_b$ of $A_{3333}$ (17.77 eV) is significantly higher compared to other configurations. Although $Er$ of $A_{3333}$ (−9.53 eV) is higher than that of $A_{2222}$ (−8.80 eV) and $A_{1122}$ (−6.90 eV), its overall stability is compromised due to its substantial $E_b$. Moreover, the $Ce^{4+}$ ions directly above the oxygen vacancy in $A_{3333}$ exhibit a stronger relaxation (0.43 Å) compared to those in $A_{2222}$ (0.23 Å) and $A_{1122}$ (0.14 Å). This indicates that while surface $Ce^{4+}$ relaxation contributes to stability, the large bond-breaking energy in $A_{3333}$ reduces its overall stability compared to $A_{2233}$. Therefore, based on this analysis, at very low $V_O$'s concentration, $V_O$'s tend to prefer the subsurface layer. As the concentration of $V_O$'s increases (from 1/16 to 1/4), the repulsion between $V_O$'s become significant, causing some $V_O$'s to migrate to the 3$^{rd}$ layer where more relaxation space is available. However, due to their high bond breaking energy associated with $V_O$'s in the 3$^{rd}$ layer, not all of $V_O$'s will migrate there. Note that already at $\Theta = 3/16$, vacancies begin populating the 3$^{rd}$ layer (**Supplementary Table 5** and **Supplementary Figure 9**). We also consider additional configurations in which the positions of $Ce^{3+}$ remain unchanged with respect to those in $A_{2233}$, $A_{2222}$ and $A_{1122}$, but those of the vacancies not. They are either distributed in the surface and subsurface ($A_{1122a}$) or scattered across the surface, subsurface, and third oxygen layer ($A_{1223}$) (**Supplementary Figure 10**). In these cases, the formation energy is higher than that of the $A_{2233}$ configuration, and the relaxation is smaller (**Supplementary Table 5**). To rule out the migration of $V_O$'s to the fourth oxygen layer, we performed DFT+U calculations for some configurations with vacancies distributed in the fourth layer (**Supplementary Figure 11**), and no configurations with lower formation energies were found. This further supports the stability of the $A_{2233}$ configuration.

**Monte Carlo sampling in a larger supercell based on machine learning model**

Furthermore, we performed simulations using a periodic slab with a (8×8) surface supercell of four O-Ce-O layers to sample the distribution of Vo's at different temperatures (T=300 K, 500 K, 700 K, 900 K, and 1100 K) and concentrations ($\Theta$=1/16, 1/8, 3/16, and 1/4) using the metropolis Monte Carlo (MC) algorithm. The energy of each configuration was predicted by the machine learning model. Each sampling process involved no less than 5 million steps, and the results are shown in **Fig. 4, Fig. 5, and Supplementary Figure 12**. The configurations presented in Fig.4 are the final snapshots obtained from each trajectory, while distinct simulated configurations from MC sampling are supplemented in **Supplementary Figure 12**, including five snapshots extracted from the entire trajectory at regular intervals.

In these simulations, configurations featuring the aggregation of local patterns formed by $Ce^{3+}$ ions and $V_O$'s, in which $V_O$'s are distributed in the subsurface and 3$^{rd}$ layers ($V_{O2}V_{O3}\_5N$) were observed. This implies that $V_O$'s prefer locating in the 3$^{rd}$ layer rather than the surface layer. Specifically, for $\Theta = 1/16$, $V_O$'s aggregate and form linear configuration of third-neighbor vacancy pairs in subsurface at 300 K, consistent with DFT predictions [25] and AFM observations [21]. As temperature increases to 500 K, $V_O$'s can migrate to the third oxygen layer and form $V_{O2}V_{O3}\_5N$ clusters with subsurface $V_O$'s. With rising temperatures, the $V_O$'s distribution becomes disordered, as reported in previous studies[32]. At 500 K and 1100 K, we observe the presence of surface $V_O$'s (**Supplementary Figure 12**). For $\Theta = 1/8$, $V_O$'s aggregate and form similar patterns as those at $\Theta =1/16$, and from 900K, we also observed surface $V_O$'s (**Fig. 4 and Supplementary Figure 12**). As the $V_O$ concentration further increases ($\Theta=3/16, 1/4$), third layer $V_O$'s begin to appear already from 300 K, and corresponding $V_{O2}V_{O3}\_5N$ motifs and third-neighbor vacancy pairs co-exist in the subsurface. Among these, for Θ=1/4 at 300 K, we also observe a local 2×2 ordered pattern in the subsurface, as previously predicted [25]. At these high vacancy concentrations, as the temperature increases, some nearest-neighbor linear $V_O$ patterns appear in the subsurface accompanied with few Vo's in the third layer. Surface Vo's appear very occasionally (**Supplementary Figure12**).

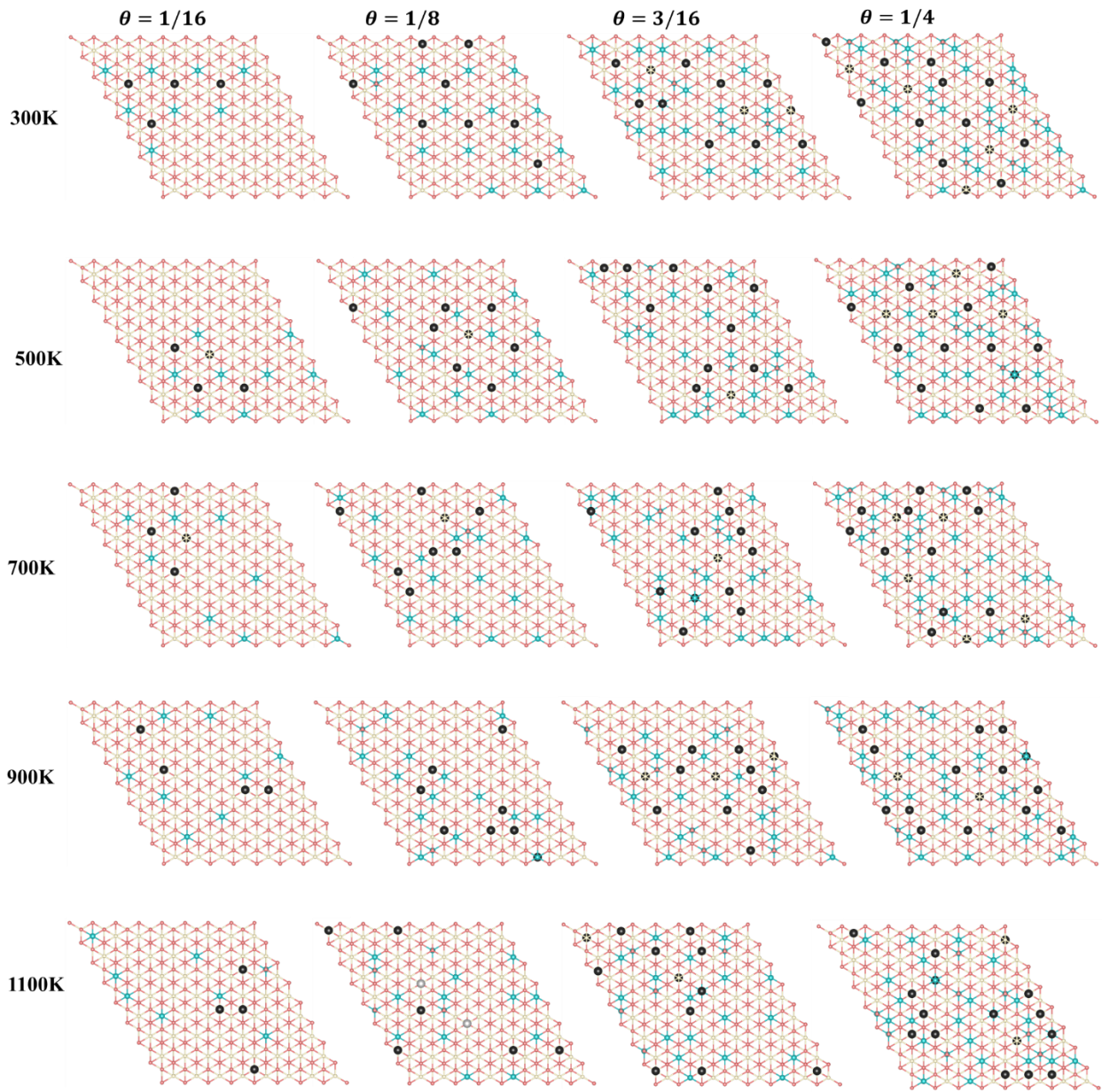

**Fig. 4 Simulated configurations after 5 million MC steps.** Top view of the final snapshots obtained from each trajectory of a (8×8) supercell at various temperatures (300 K, 500 K, 700 K, 900 K, 1100 K) and different vacancy concentrations (1/16, 1/8, 3/16, 1/4) sampled using MC methods. Structures are depicted in the unrelaxed state.

According to the statistics of the MC trajectory, with an increase in $V_O$ concentration and temperature, $V_O$'s are more likely to be distributed in the third layer than the surface. This can be understood since most of the excess electrons begin to localize at the first cerium layer, and as the concentration and temperature increase, some excess electrons begin to be located in the second cerium layer, being able to occupy second neighboring sites of $V_O$'s in the third oxygen layer (**Fig.5**).

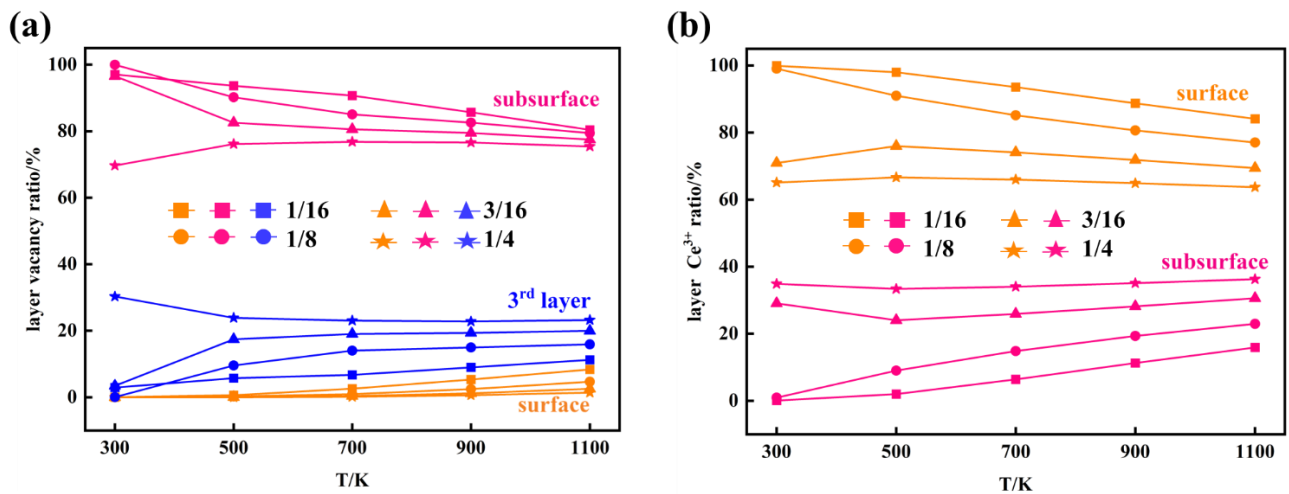

**Fig. 5 Distribution of $V_O$'s and $Ce^{3+}$'s at different temperatures and concentrations.** The distribution ratio of **(a)** $V_O$'s in the surface, the subsurface and the third layer and **(b)** The distribution ratio of $Ce^{3+}$'s in the surface, the subsurface under different $V_O$'s concentration (Θ=1/16, 1/8, 3/16, 1/4) and different temperatures (T=300 K, 500 K, 700 K, 900 K,1100 K) conditions.

We also note that an earlier STM experiment observed the abundance of surface $V_O$'s at moderate to high temperature[20]. These experiments cannot characterize the $V_O$'s beyond the subsurface. The novel finding of the relative higher stability of the 3rd layer $V_O$'s compared to surface $V_O$'s at the high $V_O$'s concentrations underscores the potential requirement to reconsider previous interpretations that neglect-ed the possible existence of vacancies in the third oxygen layer[32]. More sophisicated studies are needed in the future.

In summary, the reduced $CeO_2$ surface is a complex system featuring oxygen vacancies, $V_O$'s, and $Ce^{3+}$ ions with many-body interactions among them. Understanding these interactions is challenging without decoupling them into simpler one- and two-body terms. Through the application of the LASSO regression method and cluster expansion, we successfully achieved this decoupling with high accuracy. Among the total 3144 clusters representing many-body interactions, we identified 15 key clusters that play a predominant role in the configuration space. Our approach effectively addresses the challenge of sampling a large number of configurations and disentangling the many-body interactions among the $Ce^{3+}$ ions and $V_O$'s. Our investigation revealed a tendency for $V_O$'s on the reduced $CeO_2$(111) surface to aggregate into specific configurations with $Ce^{3+}$ ions, favoring the $V_O$'s positioning in the third oxygen layer. This phenomenon primarily stems from the significant lattice relaxation effect. This research reshapes the fundamental understanding of ceria surface structures and lays a solid foundation for further active learning when extending the approach to other Ce-related oxides. Moreover, the coupled machine-learning and cluster expansion method demonstrated here can be readily adapted to study other metal oxides with well-defined lattice structures.

**METHODS**

We performed spin-polarized DFT calculations using the Perdew-Burke-Ernzerh (PBE) functional, implemented in the Vienna *ab initio* simulation package (VASP) [38,39]. To account for the localized Ce 4f states ($Ce^{3+}$), we applied DFT+U with an effective U value of 5.0 eV [40-42]. Valence states were considered for Ce (5s, 5p, 6s, 4f, 5d) and O (2s, 2p) electrons, and projector augmented wave (PAW) potentials were employed to represent core-valence interactions. A plane-wave cutoff of 400 eV was used to expand the Kohn-Sham valence states [43,44]. The $CeO_2$ (111) surface was modeled ad a periodic slab with a (4×4) surface supercell and included four O-Ce-O tri-layers. To avoid interactions between periodic images, a vacuum layer of 15 Å was included. The bottom three atomic layers (O-Ce-O) were held fixed to their bulk positions, and the Γ point was adopted for all calculations.

Machine Learning (ML) in conjunction with the CELL code (available at the Python Package Index (PyPI) repository https://pypi.org/project/clusterX/; documentation can be found at https://sol.physik.hu-berlin.de/cell) was employed to train energy data calculated using DFT and optimize the cluster-expansion (CE) model [32,45-49]. The cluster pool consists of 3144 clusters, each containing no more than 4 points (where a point represents either a $Ce^{3+}$ or a Vo), with the distance between the points determined according to the lattice constant of 15.45 Å in the *x* and *y* directions and periodicity. After comparison, a largest distance between points (lattice sites) of 9 Å was found to cover all possible cluster types; however, no additional configurations with lower energy were found compared to a cutoff distance of 8 Å (**Supplementary Table 6**). Therefore, the largest distance between points in the clusters was set to 8 Å. Various configurations were considered, incorporating one to four vacancies randomly located at the surface, subsurface, and third oxygen layer, each corresponding to different concentrations (Θ=1/16, 1/8, 3/16, 1/4). The $Ce^{3+}$ positions were randomly located at the surface and subsurface. The Vo formation energy ($E_f^{CE}$) for a given configuration S is described as $E_f^{CE} = \Sigma_{a=1}^{N_a} m_a J_a X_{sa}$, where the sum is taken over all symmetrically inequivalent interactions[32,50]. Configuration S, encompasses various clusters, with the multiplicity of cluster $a$ represented by $m_a$. The probability of finding cluster $a$ in configuration S is expressed by $X_{sa}$ and $J_a$ represents the effective cluster interactions, for more details refer to Ref. 36. During the training process, a binary-linear cluster basis was set. For the selection of relevant clusters and fitting of model, the LASSO algorithm was chosen, with the maximal sparsity parameter set to 0.1, and minimal sparsity parameter set to 0.000001, and alpha set to 0.000294. Additionally, the Leave-One-Out-Cross-Validation method was employed to judge the predictability of the model and avoid overfitting.

ACKNOWLEDGEMENT

Y. G. thanks the support of the National Key R&D Program of China (2023YFA1506903), National Natural Science Foundation of China (12174408), Natural Science Foundation of Shanghai (22JC1404200), and the Foundation of Key Laboratory of Low-Carbon Conversion Science & Engineering, Shanghai Advanced Research Institute, Chinese Academy of Sciences (KLLCCSE-202201Z, SARI, CAS). H. L. thanks the funding support from National Key R&D Program of China (2022YFA1504001) and the National Natural Science Foundation of China (22288102, 21935001). M.V.G.P. thanks the support of the Grant PID2021-128915NB-I00 funded by MCIN/AEI/10.13039/501100011033 and, by "ERDF A way of making Europe". All calculations were performed at National Supercomputer Centers in Tianjin, Shanghai and Guangzhou.

COMPETING INTERESTS

The Authors declare no Competing Financial or Non-Financial Interests.

DATA AVAILABILITY

The CELL code can be available via at the Python Package Index (PyPI) repository https://pypi.org/project/clusterX/. Data of machine learning can be found at DOI：10.17172/NOMAD/2024.05.08-1.

## AUTHOR CONTRIBUTIONS

Y. G. conceived the original idea. H. L., M. V. G. P., and Y. G. supervised the project. Y. Z. performed the DFT calculations, Y. Z., Z.-K. H., and X. H. performed the ML and CE calculations. Y. Z. analyzed the data. Z.-K. H., and X. H. helped to analyze the data. M. T., S. R., and C. D. developed the program code. Y. Z. wrote the initial draft of the manuscript. M. V. G. P. and Y. G. assisted in interpreting the results and revising the manuscript. All authors contributed to discussions and provided critical feedback.

## REFERENCES

1. Rodriguez, J. A. et al. Activity of $CeO_X$ and $TiO_X$ Nanoparticles Grown on Au (111) in the Water-Gas Shift Reaction. *Science* **318**, 1757-1760 (2007).

2. Park, S., Vohs, J.M. and Gorte, R.J. Direct oxidation of hydrocarbons in a solid-oxide fuel cell. *Nature* **404**, 265-267 (2000).

3. Fu, Q., Saltsburg, H., and Flytzani-Stephanopoulos, M. Active Nonmetallic Au and Pt Species on Ceria-Based Water-Gas Shift Catalysts. *Science* **301**, 935-938 (2003).

4. Deluga, G. A., Salge, J. R., Schmidt, L.D. and Verykios, X. E. Renewable Hydrogen from Ethanol by Autothermal Reforming. *Science* **303**, 993-997 (2004).

5. Campbell, C.T., and Peden, C.H.F. Oxygen Vacancies and Catalysis on Ceria Surfaces. *Science* **309**, 713-714 (2005).

6. Wang, X. Q. et al. In Situ Studies of the Active Sites for the Water Gas Shift Reaction over $Cu-CeO_2$ Catalysts: Complex Interaction between Metallic Copper and Oxygen Vacancies of Ceria. *J. Phys. Chem. B* **110**, 428-434 (2006).

7. Gorte, R. J. Ceria in Catalysis: From Automotive Applications to the Water–Gas Shift Reaction. *AIChE Journal* **56,** 1126-1135 (2010).

8. Rodriguez, J. A., Liu, P., Hrbek, J., Evans, J., and Pérez, M. Water Gas Shift Reaction on Cu and Au Nanoparticles Supported on $CeO_2(111)$ and ZnO ($000\bar{1}$): Intrinsic Activity and Importance of Support Interactions. *Angew. Chem. Int. Ed*. **46**, 1329-1332 (2007).

9. Bruix, A. et al. A New Type of Strong Metal−Support Interaction and the Production of $H_2$ through the Transformation of Water on $Pt/CeO_2(111)$ and $Pt/CeOx/TiO2(110)$ Catalysts. *J. Am. Chem. Soc*.**134**, 8968-8974 (2012).

10. Capdevila-Cortada, M., García-Melchor, M. and López, N. Unraveling the structure sensitivity in methanol conversion on $CeO_2$: A DFT + U study. *Journal of Catalysis* **327**, 58-64 (2015).

11. Nörenberg, H. and Briggs, G.A.D. Surface Science Letters Defect formation on $CeO_2$ (111) surfaces after annealing studied by STM. *Surface Science* **424**, L352-L355 (1999).

12. Fukui, K. -I., Namai, Y. and Iwasawa,Y. Imaging of surface oxygen atoms and their defect structures on $CeO_2$(1 1 1) by noncontact atomic force microscopy. *Applied Surface Science* **188**, 252-256 (2002).

13. Olbrich, R. et al. Surface Stabilizes Ceria in Unexpected Stoichiometry. *J. Phys. Chem. C* **121**,6844-6851 (2017).

14. Gritschneder, S. and Reichling, M. Structural elements of CeO2 (111) surfaces. *Nanotechnology* **18** 044024 (2007).

15. Gritschneder, S., Namai, Y., Iwasawa, Y. and Reichling, M. Structural features of $CeO_2$ (111) revealed by dynamic SFM. *Nanotechnology* **16**, S41-S48 (2005).

16. Torbrügge, S. Cranney, M. and Reichling, M. Morphology of step structures on $CeO_2$(111). *Appl. Phys. Lett*. **93**, 073112 (2008).

17. Šutara, F. et al. Epitaxial growth of continuous $CeO_2$(111) ultra-thin films on Cu(111). *ScienceDirect* **516**, 6120-6124 (2008).

18. Škoda, M. et al. Interaction of Au with: A photoemission study. *J. Chem. Phys*. **130**, 034703(2009).

19. Grinter, D.C. and Thornton, G. Structure and reactivity of model $CeO_2$ surfaces. *J. Phys.: Condens. Matter* **34**, 253001(2022).

20. Esch, F. et al. Electron Localization Determines Defect Formation on Ceria Substrates. *Science* **309**, 752-755 (2005).

21. Torbrügge, S., Reichling, M., Ishiyama, A., Morita, S. and Custance, Ó. Evidence of Subsurface Oxygen Vacancy Ordering on Reduced $CeO_2$(111). *Phys. Rev. Lett*. **99**, 056101 (2007).

22. Skorodumova, N. V., Simak, S. I., Lundqvist, B. I., I. Abrikosov, A. and Johansson, B. Quantum Origin of the Oxygen Storage Capability of Ceria. *Phys. Rev. Lett*. **89**, 166601 (2002).

23. Li, H.-Y. et al. Multiple configurations of the two excess 4f electrons on defective CeO2(111): Origin and implications. *Phys. Rev. B* **79**, 193401 (2009).

24. Ganduglia-Pirovano, M. V., Silva, J. L. F. D. and Sauer, J. Density-Functional Calculations of the Structure of Near-Surface Oxygen Vacancies and Electron Localization on CeO$_2$(111). *Phys. Rev. Lett*. **102**, 026101 (2009).

25. Murgida, G. E. and Ganduglia-Pirovano, M. V. Evidence for Subsurface Ordering of Oxygen Vacancies on the Reduced CeO$_2$(111) Surface Using Density-Functional and Statistical Calculations. *Phys. Rev. Lett*. **110**, 246101 (2013).

26. Kullgren, J. et al. Oxygen Vacancies versus Fluorine at CeO$_2$(111): A Case of Mistaken Identity? *Phys. Rev. Lett*. **112**, 156102 (2014).

27. Wu, X.-P. and Gong, X.-Q. Clustering of Oxygen Vacancies at CeO$_2$(111): Critical Role of Hydroxyls. Gong, *Phys. Rev. Lett*. **116**, 086102 (2016).

28. Wolf, M. J., Kullgren, J. and Hermansson, K. Comment on "Clustering of Oxygen Vacancies at CeO$_2$(111): Critical Role of Hydroxyls". *Phys. Rev. Lett*. **117**, 279601 (2016).

29. Wu, X.-P. and Gong, X.-Q. Wu and Gong Reply, *Phys. Rev. Lett*. **117**, 279602 (2016).

30. Kullgren, J., Wolf, M. J., Mitev, P. D., Hermansson, K. and Briels, W. J. DFT-based Monte Carlo Simulations of Impurity Clustering at CeO$_2$(111). *J. Phys. Chem. C* **121**, 15127-15134 (2017).

31. Jerratsch, J.-F. et al. Electron Localization in Defective Ceria Films: A Study with Scanning-Tunneling Microscopy and Density-Functional Theory. *Phys. Rev. Lett*. **106**, 246801 (2011).

32. Han, Z.-K., Yang, Y.-Z., Zhu, B., Ganduglia-Pirovano, M. V. and Gao, Y. Unraveling the oxygen vacancy structures at the reduced CeO$_2$(111) surface. *Phys. Rev. Materials*. **2**, 035802 (2018).

33. Murgida, G. E., Ferrari, V., Llois, A. M. and Ganduglia-Pirovano, M. V., Reduced CeO$_2$(111) ordered phases as bulk terminations: Introducing the structure of Ce$_3$O$_5$. *Phys. Rev. Mat*. **2**, 083609 (2018).

34. Murgida, G. E., Ferrari, V., Ganduglia-Pirovano, M. V., and Llois, A. M. Ordering of oxygen vacancies and excess charge localization in bulk ceria: A DFT+U study. *Phys. Rev. B*. **90**, 115120 (2014).

35. Tibshirani, R. Regression Shrinkage and Selection Via the Lasso. *J. R. Statist. Soc.: Ser. B* (Methodological) 58, 267-288 (1996).

36. Sarker, D., Han, Z.-K. and Levchenko, S. V. Iterative cluster expansion approach for predicting the structure evolution of mixed Ruddelsden-Popper oxides $La_{2-x}Sr_xNi_{1-y}Fe_yO_{4\pm\delta}$. *Phys. Rev. Materials*. 7, 055802 (2023).

37. Barroso-Luque, L. et al. Cluster expansions of multicomponent ionic materials: Formalism and methodology. *Phys. Rev. B* 106, 144202 (2022).

38. Kresse, G. and Furthmüller, J. Efficient iterative schemes for ab initio total-energy calculations using a plane-wave basis set. *Phys. Rev. B* **54**, 11169-11186 (1996).

39. Perdew, J. P., Burke, K. and Ernzerhof, M. Phys. Generalized Gradient Approximation Made Simple. *Rev. Lett*. **77**, 3865-3868 (1996).

40. Dudarev, S. L., Botton, G.A., Savrasov, S. Y., Humphreys, C. J. and Sutton, A. P. Electron-energy-loss spectra and the structural stability of nickel oxide: An LSDA+U study. *Phys. Rev. B* **57**, 1505-1509 (1998).

41. Cococcioni, M. and Gironcoli, S. D. Linear response approach to the calculation of the effective interaction parameters in the LDA+U method. *Phys. Rev. B* **71**, 035105 (2005).

42. Castleton, C. W. M., Kullgren, J. and Hermansson, K. Tuning LDA+U for electron localization and structure at oxygen vacancies in ceria. *J. Chem. Phys*. **127**, 244704 (2007).

43. Blöchl, P. E. Projector augmented-wave method. *Phys. Rev. B* **50**, 17953-17979 (1994).

44. Kresse, G. and Joubert, D. From ultrasoft pseudopotentials to the projector augmented-wave method. *Physical Review B* **59**, 1758 -1775 (1999).

45. Cell documentation, https://sol.physik.hu-berlin.de/cell.

46. Rigamonti, S., Troppenz, M., Kuban, M., Hübner, A. and Draxl, C. https://arxiv.org/abs/2310.18223 (2023).

47. Nguyen, M. C., Zhao, X., Wang, C.-Z. and Ho, K.-M. Cluster expansion modeling and Monte Carlo simulation of alnico 5-7 permanent magnets. *J. Appl. Phys*. **117**, 093905 (2015).

48. Troppenz, M., Rigamonti, S. and Draxl, C. Predicting Ground-State Configurations and Electronic Properties of the Thermoelectric Clathrates $Ba_8Al_xSi_{46-x}$ and $Sr_8Al_xSi_{46-x}$. *Chem. Mater*. **29**, 2414-2424 (2017).

49. Han, Z.-K. et al. First-principles study of Pd-alloyed Cu (111) surface in hydrogen atmosphere at realistic temperatures. *J. Appl. Phys*. **128**, 145302 (2020).

50. Sanchez, J. M., Ducastelle, F. and Gratias, D. Generalized cluster description of multicomponent systems. *Physica A: Statistical Mechanics and its Applications*. **128**, 334-350 (1984).

# Figures legends

**Scheme 1. Machine learning flowchart of CE model.**

**Fig.1. Model evaluation and primary features from Machine-Learning.** The error between the cluster-expansion model predicted energy and the DFT energy for **(a)** training set and **(b)** leave-one-out cross-validation. Violin plots of the absolute error in the energy of configurations for **(c)** training set and **(d)** leave-one-out cross-validation. The upper and lower limits of the rectangles represent the 75th and 25th percentiles of the distribution, the internal white dot marks the median (50th percentile), and the upper and lower limits of the thin line extending from the rectangle bars indicate the minimum and maximum errors. The violin area represents the probability of data distribution. **(e)** 15 primary features predicted by machine learning and their energies, with structures depicted in the unrelaxed state. $Ce^{4+}$, $Ce^{3+}$, oxygen atoms, and surface oxygen vacancy are depicted in white, blue, pink, and light gray respectively. Subsurface and third oxygen layer vacancy are depicted in black.

**Fig.2. Predicted relative energies of selected motifs of Vo's and $Ce^{3+}$'s based on the identified 15 primary features.** **(a)-(e)** $Ce^{3+}$ ion is located in the first cerium layer, **(f)-(j)** $Ce^{3+}$ ion is located in the second cerium layer. $Ce^{4+}$, $Ce^{3+}$, oxygen atoms, and surface oxygen vacancy are depicted in white, blue, pink, and light gray respectively, subsurface and third oxygen layer vacancy are depicted in black. Structures are depicted in the unrelaxed state.

**Fig.3. Top views, side views, and the Vo's formation energies of the configurations $A_{2233}$, $A_{2222}$, $A_{1122}$ and $A_{3333}$.** **(a)** configuration $A_{2233}$, with two $V_O$'s in the subsurface layer and two $V_O$'s in the 3$^{rd}$ oxygen layer, **(b)** configuration $A_{2222}$, with four $V_O$'s in the subsurface layer, **(c)** configuration $A_{1122}$, with two $V_O$'s in the subsurface layer and two $V_O$'s in the surface layer. **(d)** configuration $A_{3333}$, with four $V_O$'s in the 3$^{rd}$ layer. $Ce^{4+}$, $Ce^{3+}$, oxygen atoms and surface oxygen vacancy are depicted in white, blue, pink, and light gray respectively, subsurface and third oxygen layer vacancy are depicted in black, purple atoms are $Ce^{4+}$. Top and side views are depicted in the unrelaxed and relaxed state, respectively. Arrows indicate the relaxation direction of the purple-colored $Ce^{4+}$ at the surface layer.

**Fig.4. Simulated configurations after 5 million MC steps.** Top view of the final snapshots obtained from each trajectory of a (8×8) supercell at various temperatures (300 K, 500 K, 700 K, 900 K, 1100 K) and different vacancy concentrations (1/16, 1/8, 3/16, 1/4) sampled using MC methods. Structures are depicted in the unrelaxed state.

**Fig.5. Distribution of $V_O$'s and $Ce^{3+}$'s at different temperatures and concentrations.** The distribution ratio of **(a)** $V_O$'s in the surface, the subsurface and the third layer and **(b)** The distribution ratio of $Ce^{3+}$'s in the surface, the subsurface under different $V_O$'s concentration ($\Theta$=1/16, 1/8, 3/16, 1/4) and different temperatures (T=300 K, 500 K, 700 K, 900 K, 1100 K) conditions.